\documentclass[12pt]{article}
\usepackage{epsfig}

\begin{document}
\begin {center}
{\bf \Large Experimental disagreements with Extended Unitarity}
\vskip 5mm
{D.\ V.\ Bugg\footnote{email: D.Bugg@rl.ac.uk} \\
{\normalsize\it Queen Mary, University of London, London E1\,4NS, UK}
\\[3mm]}
\end {center}
\date{\today}

\begin{abstract}
In production processes, e.g. $J/\Psi \to \omega \pi \pi$ or
$\bar pp \to 3\pi$, the $\sigma$ and $f_0(980)$ overlap in the
same partial wave.
The conjecture of Extended Unitarity (EU) states that the
$\pi\pi$ pair should have the same phase variation as
$\pi\pi$ elastic scattering.
This is an extension of Watson's theorem beyond its original
derivation, which stated only that the $s$-dependence of a
single resonance should be universal.
The prediction of EU is that the deep dip observed in
$\pi \pi$ elastic scattering close to 1 GeV should also
appear in production data.
Four sets of data disagree with this prediction.
All require different relative magnitudes of
$\sigma$ and $f_0(980)$.
That being so, a fresh conjecture is to rewrite the 2-body
unitarity relation for production  in terms of observed
magnitudes.
This leads to a prediction different to EU.
Central production data from the AFS experiment fit
naturally to this hypothesis.

\vskip 2mm

{\small PACS numbers: 13.25.-k, 13.25.Gv, 13.75.Lb}
\end{abstract}
\section {Introduction}
In its simplest form, the idea of
Extended Unitarity (EU) states that the $\pi \pi$ pair
in a single partial wave should have the same phase variation
with $s$ in {\it all} reactions as in elastic scattering.
This idea originates from Aitchison \cite {Aitchison} and has
been adopted in various guises by many authors.
His arguments will be presented in detail in Section 2,
so as to expose the assumptions and consequences.
At the time the idea was introduced, it was a reasonable
conjecture; now modern data allow it to be checked accurately,
but disagree with it.

Many experimental groups have made extensive fits to production
data using a $K$-matrix approach based on EU.
These fits are excellent; no criticism is intended of their
quality.
Experimentalists have found empirically the necessary freedom
to get good fits to data.
However, on close inspection, this freedom is inconsistent
with strict EU.

 This whole topic has been the subject of extensive discussion
with many authors.
There is a bewildering jungle of claims and counter-claims.
My objective is to cut a path through this tangle and
expose where problems lie; this makes the presentation
pedantic in places.

Aitchison's essential point is that all processes should be
described by a universal denominator
$[1 - i\rho(s)K(s)]$, where $K$
is the same as for elastic scattering; $\rho$ is Lorentz
invariant phase space.
The assumption which is being made is that Watson's theorem
\cite {Watson} applies to the coherent sum of all components
in the  $J^P = 0^+$ partial wave.
This is a step beyond Watson's derivation, which referred
only to a single eigenstate; Watson did not consider
overlapping resonances.

\begin{figure}[htb]
\begin{center}
\vskip -12mm
\epsfig{file=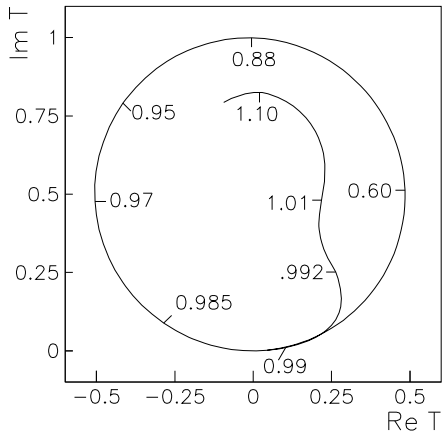,width=7cm}
\vskip -7mm
\caption {Argand plot of the $\pi \pi$ $I=0$
S-wave in elastic scattering; masses are marked in GeV.}
\end{center}
\end{figure}

In $\pi \pi$ elastic scattering, the $f_0(980)$ is
superimposed on a slowly rising amplitude associated with the
$\sigma$ pole.
Cern-Munich data \cite {Hyams} show that the phases of
these two components add.
Below the $KK$ threshold, both $\sigma$ and $f_0(980)$
$T$-matrices $T=e^{i\delta}\sin \delta$ are confined to the
unitarity circle if we neglect the tiny inelasticity due to
$\pi \pi \to \gamma \gamma$.
Unitarity may be satisfied by multiplying $S$-matrices
$S = e^{2i\delta}$ of $\sigma$ and $f_0(980)$,
as suggested by Dalitz and Tuan \cite {Tuan}.
This fits the data within errors of $\sim 3.5^\circ$.

Fig. 1 shows the Argand diagram for the $I=0$ $\pi \pi$
S-wave from my recent re-analysis of these (and
other) data \cite {f01370}.
From BES data on $J/\Psi \to \phi \pi ^+ \pi ^-$, the
$f_0(980)$ has a full-width at half-maximum of $34 \pm 8$ MeV
\cite {phipp}.
The combined phase shift rises rapidly from $90^\circ$
at 0.88 GeV to $270^\circ$ near 1.1 GeV.
There is a deep dip in the cross section where the combined
phase goes rapidly through $180^\circ$.
The crucial point of EU is that this feature should be common
to production processes.

\begin{figure}[htb]
\begin {center}
\vskip -15mm
\epsfig{file=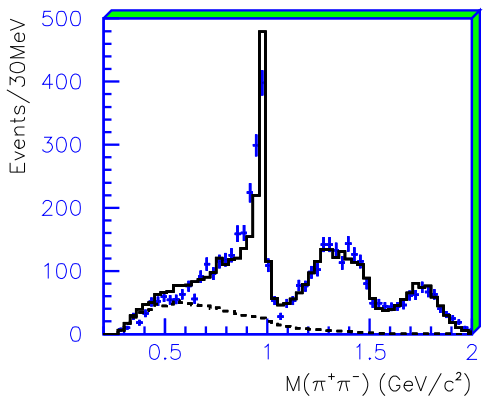,width=7cm}
\vskip -8mm
\caption {The $\pi \pi$ mass projection for BES data on
$J/\Psi \to \phi \pi ^+\pi ^-$:
the upper histogram shows the current fit to experimental
points [5];
the lower histogram shows the fitted $\sigma$ component.}
\end {center}
\end{figure}

BES data on $J/\Psi \to \phi \pi \pi$  \cite {phipp}
immediately require a modification of the rudimentary form of
EU.
The $\pi \pi$ mass spectrum in these data is reproduced in
Fig. 2.
There is a dominant $f_0(980)$ contribution and a small
interfering $\sigma$ contribution; this is very different to
elastic scattering.
L\" ahde and Meissner \cite {Lahde} modify the conjecture of
EU to apply separately to strange and non-strange components,
i.e. to the scalar form factors for $\pi \pi$ and $KK$.

The dip in the elastic cross section at 989 MeV is a very
delicate feature.
If, for any reason, relative magnitudes of $\sigma$ and
$f_0$ change, the zero at 989 MeV can disappear quickly;
here and elsewhere, $f_0$ will denote $f_0(980)$ unless there
is confusion with other $f_0$'s.
If the phase of $f_0$ changes with respect to the
$\sigma$, the mass at which the dip appears will likewise
change.
The interference region between $\sigma$ and $f_0(980)$
is an ideal place to check Extended Unitarity.

Two considerations will play a critical role: unitarity and
analyticity.
Unitarity is often quoted and plays an essential role in
setting up the current $K$-matrix formalism which treats both
elastic scattering and production on the same basis.
For a production reaction, Aitchison conjectures a unitarity
relation for the production amplitude $F$:
\begin {equation}
{\rm Im}\, F = FT_{el}^*.
\end {equation}
He defines his $F$ to be proportional to $T^{(p)}/\rho$,
where $T^{(p)}$ refers to production:
\begin {equation}
F\rho  = \alpha T^{(p)}.
\end {equation}
Dividing both sides of (1) by $\alpha$,
\begin {equation}
{\rm Im}\, T^{(p)} = T^{(p)}T^*_{el}.
\end {equation}
It is odd that $T^{(p)}$ on the right-hand side is
multiplied by $T^*_{el}$, unless $T^{(p)} = T_{el}$.
However, experiment will require different contributions to
$T^{(p)}$ from $\sigma$ and $f_0$.

Consider next analyticity.
Dispersion relations connect magnitudes and phases.
If the relative magnitudes of $f_0$ and $\sigma$ change
from those of elastic scattering because of matrix elements,
their relative phases {\it must} also change.
Conversely, analyticity predicts that if the phase
variation with $s$ of the amplitude is universal, as EU
demands, so is  the variation with $s$ of the magnitude (up to a
constant scaling factor);
for the simplest situation where only resonances are present,
the relative magnitudes of $\sigma$, $f_0(980)$ and any
further $f_0$ must be {\it almost} the same in production as elastic
scattering.
This is a point which has almost always been ignored.

The word `almost' represents a caveat: there may in addition be a
polynomial is $s$ which can be different between elastic
scattering and production.
It turns out that one can plausibly limit deviations of relative
magnitudes within $12\%$.
This question is discussed in subsection  2.1.
Experiment requires larger deviations than this in
the four sets of data discussed here.
This implies phases must change from those predicted by
EU.
Experimentalists have correctly allowed for this by using
complex coupling constants in the isobar model.

Section 3 compares the prediction of EU with 3 sets of
data.
The first concerns BES data for $J/\Psi \to \omega \pi^+\pi ^-$
\cite {WPP}.
The non-strange components of $\sigma$ and $f_0$ dominate both
elastic scattering and $J/\Psi \to \omega \pi \pi$.
From Aitchison's algebra and that of L\"{a}hde and Meissner, it
follows that the $f_0$ amplitude should have almost the
same magnitude as the $\sigma$ amplitude, as well as the same
phase as elastic scattering.
This prediction is contradicted by the data, where no $f_0(980)$
is visible and a fit to the data places a low limit on it.

The next two sets are Crystal Barrel data for $\bar pp \to 3\pi
^0$, where $\sigma$ and $f_0(980)$ are clearly visible
\cite {CBAR}.
One set is for annihilation in liquid hydrogen and the other for
gaseous hydrogen.
Annihilation from the $^3P_1$ initial state is 13\% in liquid
and 48\% in gas, allowing a clear separation of amplitudes for
production of $\sigma$ and $f_0$ from $^1S_0$ and $^3P_1$.
Results for both are inconsistent with the deep dip of elastic
scattering predicted by EU.

Section 4 concerns data from the AFS experiment on central production:
$pp \to pp\pi^+\pi^-$ \cite {AFS}.
Here one expects the protons in the final state to act as spectators.
However, EU still fails conspicuously to fit the data.
This important result leads to a revised form of the unitarity
relation, as follows.

\begin{figure}[htb]
\begin{center}
\vskip -87mm
\epsfig{file=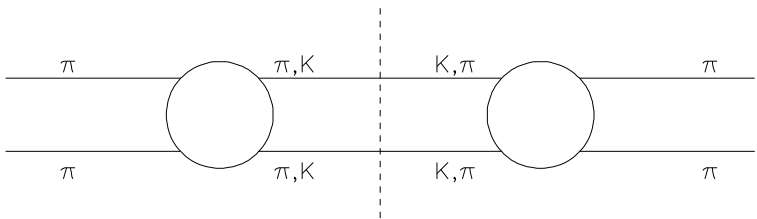,width=16cm}
\vskip -52mm
\end {center}
\caption {Unitarity diagram for $\pi \pi \to \pi \pi$.}
\end{figure}

Fig. 3 sketches the usual diagrammatic approach to
the unitarity relation.
It may be derived by cutting the diagram down the middle, along the dashed
line.
For a 2-body system of $\pi \pi$, $KK$, $\eta \eta$, etc.
the resulting relation is well known:
\begin {equation}
{\rm Im}\, T_{el} = T_{el}T_{el}^*.
\end {equation}
The application of 2-body unitarity assumes that the pions interact
only with one another, not with any spectator.
In most sets of data there are large signals where pions
do interact with the spectator.
For $J/\Psi \to \omega \pi \pi$, as an example,
the $b_1(1235)\pi$ channel accounts for 40\% of events
\cite {WPP}.
Some of it may be generated by pions from decays of $\sigma$ or
$f_0$ rescattering from the spectator; this is a so-called triangle
graph.
Aitchison himself remarks that this can distort the unitarity relation.
This provides one reason why EU may fail for the first three sets of
data;  it does not explain the fourth, where some further effect is
required.

There are fundamental differences between elastic scattering and
production.
In $J/\Psi \to \omega \pi \pi$, for example, matrix elements
$<J/\Psi|\omega \sigma>$ and $J/\Psi|\omega f_0>$ dictate the
magnitudes of these amplitudes; any values are possible.
This differs from elastic scattering, where $\sigma$ and $f_0$
magnitudes are fixed purely by their coupling constants $g_\pi$ to
$\pi \pi$.
Equation (3) is an asymmetric relation, allowing $\sigma$ and $f_0$ to
be produced with different magnitudes, but requiring that they
rescatter an in elastic scattering.
A more logical alternative is the symmetric relation
\begin {equation}
{\rm Im}\, T^{(p)} = T^{(p)}T^{(p)*},
\end {equation}
hence ${\rm Im}\, F = FT^{*(p)}$.
This relation fits AFS data for central production naturally,
whereas EU does not.
If the $f_0$ is absent from production data,
(5) reduces to the obvious relation ${\rm Im}\, T^\sigma =
|T^\sigma|^2$; Aitchison's form of the relation, taken with analyticity
does not allow the $f_0$ to be absent, as we shall see in Section 2.

Section 5 suggests a new way of fitting 2-body data.
Section 6 then summarises conclusions.

\section {The hypothesis of Extended Unitarity}
In a two-body process, the scattering of a pair of pions to final
states $\pi \pi$, $KK$, $\eta \eta$, $4\pi$ and $\gamma \gamma$
must satisfy unitarity.
The $T$-matrix for these coupled channels may be written in
terms of a real $K$-matrix as
\begin {equation}
T_{el} = \rho K(1 - i\rho K)^{-1}\;.
\end {equation}
It is normalised here so that $T_{\pi \pi} =
(\eta e^{2i\delta} - 1)/2i$.
Below the inelastic threshold
\begin {equation}
\rho K = \tan \delta.
\end {equation}
The $T$-matrices used here will include couplings to all
channels.
However, it simplifies the presentation of essential points
to reduce the formalism initially to a single $\pi \pi$
channel.
This simplication is sufficient to expose the basic issues,
and can be generalised later to include inelasticity.

The approach of Aitchison \cite {Aitchison} will now be
outlined.
I am grateful to him for clarifying the algebra in more
detail than is to be found in the original publication.
Suppose $S$-matrices multiply, i.e. phases add.
Let $K_A$ and $K_B$ be $K$-matrices for $\sigma$ and $f_0$
respectively.
The elementary expression for $\tan (\delta_A + \delta _B)$
then gives a $K$-matrix for elastic scattering
\begin {equation}
K_{el} = \frac {K_A + K_B}{1 - \rho ^2 K_A K_B},
\end {equation}
from which it follows that the $T$-matrix for elastic
scattering  is
\begin {equation}
T_{el} = \frac {(K_A + K_B)\rho}{(1 - i \rho K_A)(1 - i\rho K_B)}.
\end {equation}

Aitchison now conjectures that an amplitude $F$ for producing
a two-body channel present in $K_{el}$ may be written in terms of
a vector $P$, with
\begin {eqnarray}
F &=& (1 - i \rho K_{el})^{-1}\, P \\
P &=& \frac {\alpha_AK_A + \alpha_BK_B} {1 - \rho ^2 K_AK_B},
\end {eqnarray}
where $\alpha_A$ and $\alpha_B$ are constants for production
couplings.
With this ansatz, the relation
\begin {equation}
{\rm Im}\, F = F T^*_{el},
\end {equation}
known as Extended Unitarity, is automatically satisfied.
It is a consequence of (10) that $F$ has the same phase as
$T_{el}$.
Substituting (8) and (11) in (10) gives, in this one-channel
case
\begin {eqnarray}
F  &=& \frac {\alpha_A K_A + \alpha _B K_B}
          {(1 - i\rho K_A)(1 - i\rho K_B)} \\
\; &=& \; {\alpha _A} [T_A(1 + iT_B) + \beta  T_B(1 +
iT_A)]/\rho ,
\end {eqnarray}
where $\beta = \alpha _B/\alpha _A$.
From (13), the phase of $F$ is indeed
$\delta_A + \delta _B$, as imposed by (10).
Equation (14) will play the decisive role in comparisons
with experiment.

In (14), $T_A(1 + iT_B) =
\exp i(\delta_A + \delta_B)\sin \delta_A \cos \delta_B$ and
$T_B(1 +iT_A) = \exp i(\delta_A + \delta_B)\cos \delta_A \sin
\delta_B$.
At 989 MeV, $\delta_f = 90^\circ$ and $\delta_\sigma = 92^\circ$.
So both terms are very close to zero, regardless of the values
of $\alpha_A$ and $\alpha_B$.
This predicts that production data should have the same deep
dip at this energy as elastic scattering.

There is a further point.
In the second term, $(1 + iT_A) \simeq 0$ over a sizable mass
range.
In the first term, $T_B$ should be conspicuous, since it has a
rapid phase variation and the same peak magnitude as
$T_A$, which is itself clearly visible in all sets of data
considered here.
However, the data all require the magnitude of the $f_0(980)$
to be smaller than predicted.

The key point is that the factor $1/(1 - i\rho K_B)$ of (13)
leads directly to the factor $(1 + iT_B)$ in the first term of
(14).
The first and third sets of data will require $\beta$
of (14) to be small.
In the elastic region, the first term becomes
$$
F \simeq i\alpha _A(1 + i e^{i\delta _B} \sin \delta _B).
$$
The bizarre conclusion of EU is that the phase of the
$f_0(980)$ is present even though $\alpha_B \simeq 0$.
This is inconsistent with analyticity.
It will be shown in Section 3 that experiment disagrees
with EU even without the constraint of analyticity.
However this additional constraint makes conclusions
more definitive.

\subsection {Analyticity}
For purely elastic scattering, the Omn\` es relation
\cite {Omnes} reads [including a factor $\rho (s)$ in $N(s)$]:
\begin {eqnarray}
T_{el} (s) &=& {N(s)}/{D(s)}       \\
D (s) &=& e^{-i\delta (s)} \exp -\left[ \frac {s-4m^2_\pi}{\pi}
{\rm P}\int \frac {ds'}{s'-4m^2_\pi} \frac {\delta (s')}{s-s'}
\right],
\end {eqnarray}
where P denotes the Principal Value integral.
We shall not actually need to evaluate (16).
It plays only a conceptual role and this needs considerable
explanation.
The basic point is that $D(s)$ contains both real and imaginary
parts, so $\delta(s)$ determines both.
For elastic scattering $N(s)$ is real.
It arises from the left-hand cut, i.e. meson exchanges
between the two pions.

With inelasticity, corresponding relations may be written
in a 2-channel form.
Then $\delta (s)$ is replaced by $\phi (s)$, the angle
$T$ makes to the real axis when measured from the
origin of the Argand diagram, see Fig. 12 of subsection 4.1.

If EU is valid, the production amplitude may be written
$F = X(s)/D(s)$.
In principle $X(s)$ could be anything, depending on production
dynamics.
However, we have quite precise experimental information about it.
An extreme view is that $\alpha_A$ and $\alpha_B$ of (13) can be
arbitrary and complex.
However if they are complex this leads directly to a conflict with EU.
Equation (14) contains two parts $T_A(1 + iT_B)$ and
$T_B(1 + iT_A)$.
Substituting Breit-Wigner formulae for $T_A$ and $T_B$,
the first term becomes $g^2_A(M^2_B - s)/D_A(s)D_B(s)$.
This is real but has a specific $s$-dependence in the numerator
as well as in the denominator.
If $\alpha_A$ or $\alpha_B$ becomes complex, the numerator
becomes complex.
This then introduces a phase variation separate from $D(s)$.
The ``prediction'' of EU is distorted by this extra phase.
Only if $X(s)$ is real does EU survive in its strict form.

Many experimental groups have used the P-vector
approach using complex coupling coefficients, without realising
that this destroys the universality of the phase variation with $s$.
This is what experiment demands, so they have done the right thing.
But the use of the universal denominator $[1 - i \rho K(el)]$ is
no longer logically correct.
One might as well fit directly in terms of complex coupling constants
and individual $T$-matrices for each resonance.

\subsection {Form Factors}
There is a further fundamental point.
For elastic scattering, $N(el)$ is uniquely related to
$\rm {Im}\, D(s)$ by both unitarity and analyticity.
At first sight it appears that analyticity relates $X(s)$
in the same way to $D(s)$ in production reactions, with the
result $X(s) = \alpha N(el)$, where $\alpha$ is a constant.
This requires $\beta = 1$: if the phase of the $\pi \pi$
amplitude is universal, relative magnitudes of $\sigma$ and
$f_0$ must also be universal.

There is however a caveat.
A more fundamental form of the unitarity relation (12)
is that the discontinuity of $F$ across the elastic branch
cut is $2iFT^*$.
Then $F$ may be multiplied by a polynomial $X(s)$,
providing it does not have a discontinuity along the real
$s$-axis.
A few examples will hopefully clarify ideas.
Firstly, a form factor in $s$ is one such example, arising
from the sizes of particles, i.e. from matrix elements.
Secondly, in $\phi \to \gamma f_0$, the E1 transition has an
intensity proportional to the cube of the photon momentum;
this inflates the lower side of the $f_0(980)$.
Thirdly, in $^3P_1$ $\bar pp \to \pi ^0 \sigma$, there is an
$L=1$ centrifugal barrier for the production process.
Fourthly, in some special cases, matrix elements may go through zero
as a function of $s$.
Taking $X(s)$ to be real, let us write in general
\begin {equation}
F = X(s)N(el)/D(s).
\end {equation}

A feature of all production data considered here is a strong
low-mass $\pi \pi$ peak due to the $\sigma$ pole.
This peak is not present in elastic scattering because of
an Adler zero in the elastic scattering amplitude at
$s_A\simeq 0.41 m^2_\pi$, just below the $\pi \pi$ threshold.
The elastic amplitude rises approximately linearly with
$s$ and there is no low mass peak.
The origin of the difference has been known to theorists for at
least 20 years.
Au, Morgan and Pennington \cite {AMP} pointed out
that the difference between elastic scattering and central
production data can be accomodated by using the same
Breit-Wigner denominator for both, but replacing the numerator
$N(el)$ by something close to a constant.
This polynomial is allowed because $s_A$ is outside the physical
region.
Data require $X(s)N(el) \simeq 1$, hence $X(s) \simeq 1/N(el)
\simeq 1/(s - s_A)$.
More exactly,
\begin {equation}
X(s) = 1/[(s-s_A)(1 + bs)\exp[-(s - M_A^2)/A]
\end {equation}
for the parametrisation of the $\sigma$ amplitude in
\cite {DJphysG}.
In practice, quadatic and cubic terms in $s$ are very small and
under tight control from fits to data up to 1.8 GeV.

For $J/\Psi \to \omega \pi \pi$,
the $\sigma$ pole is visible by eye in Fig. 4(a) below.
The phase of the $\sigma$ amplitude
in this reaction is experimentally the same as in elastic scattering
within $\sim 3.5^\circ$ \cite {sigphase}.
Values of $N(prodn) = X(s) N(el)$ can be determined directly from the
data.
The same is true of  the $\kappa$ \cite {kappa}, which likewise has an
Adler zero in the numerator for elastic scattering, but not for
production.
In both cases, $N(prodn)$ is consistent within errors with a constant;
the Adler zero in the numerator of elastic scattering has disappeared.
One can try fitting the $\sigma$ and $\kappa$ poles in production data
with the conventional form factor $N(prodn) = \exp (-k^2R^2/6)$,
where $k$ is momentum in the production channel.
For both, $R^2$ optimises at slightly negative values, which
are unphysical.
For the $\sigma$, $R^2 < 0.30$ fm$^2$ with $95\%$ confidence
and for the $\kappa$, $R^2 <0.38$ fm$^2$ at the same level.

A crucial piece of information in testing EU will be relative
magnitudes of $\sigma$ and $f_0$.
The magnitude of the $\sigma$ amplitude is easily separated
from the tail of the $f_0(980)$ at  920 MeV, three half-widths
from 989 MeV.
If $X(s)$ is determined at this energy and at 400 MeV,
the $\sigma$ amplitude changes by $20\%$ at 989 MeV for
the $95\%$ confidence level quoted above. However, this
exaggerates the error, since there are compensating changes in the
line-shape fitted to the amplitude.
Realistically, changes are half this.
Adding in quadrature uncertainties in the $\sigma$ line-shape
due to the opening of the $KK$ threshold,
the uncertainty in the amplitude extrapolated from 920 to 989 MeV
is  $<12\%$ with $95\%$ confidence.
This provides a tight constraint on the  relative magnitudes of
$\sigma$ and $f_0(980)$ amplitudes if EU is  correct.
This disposes of the paradox that the phase of
$f_0(980)$ can be present with $\alpha_B = 0$.

From this point onwards, it will be assumed that EU should be
supplemented with the condition imposed by analyticity within $12\%$.

\subsection{Formulae for $\sigma$ and $f_0$}
Formulae and numerical parameters for the $\sigma$ amplitude
are to be found in Refs. \cite {DJphysG} and \cite {f01370}.
The $\pi\pi$ coupling has been fitted to four sources:
(i) phase shifts deduced from Cern-Munich data by Ochs \cite {Ochs},
(ii) $K_{e4}$ data of Pislak et al. \cite {Pislak},
(iii) predictions of $\pi \pi$ phase shifts by Caprini et al.
\cite {CCL} using the Roy equations, and
(iv) BES data on $J/\Psi \to \omega \pi ^+\pi ^-$ \cite {WPP}.

The coupling to $KK$ and $\eta \eta$ has been fitted to
available data on those channels \cite {Reconcil}, and in
\cite {f01370} the $4\pi$ coupling has been derived from
Cern-Munich data.
There is close consistency between all these sets of data.
Because Refs. \cite {DJphysG} and \cite {f01370} fit the same data
with different formulae, the amplitudes agree within errors
up to 1.2 GeV.
Those of Ref. \cite {f01370} are more cumbersome to
use, since they allow for the dispersive effect of the $4\pi$
threshold.
Therefore the first and fourth sets of data discussed below
are fitted with the formulae from Ref. \cite {DJphysG}.

The general procedure adopted here is to allow the parameters
of the $\sigma$ to have the freedom allowed in earlier
determinations of its parameters, but no extra freedom
in the vicinity of $f_0(980)$.
In testing EU,  the magnitude of $f_0$ is restricted to the
$12\%$ discussed above; its phase is allowed the freedom with
which its parameters are known.

\subsection {Parameters of $f_0(980)$}
The $f_0(980)$ is so narrow that it is readily separated
from the $\sigma$.
The BES data on $J/\Psi \to \phi \pi ^+\pi ^-$ shown in
Fig. 2 exhibit a very clear $f_0(980) \to \pi \pi$ signal.
An important point is the excellent mass resolution and mass
calibration of the BES detector, $\sim 4$ MeV.
Both are easily checked for the $KK$ channel against the
very precisely known parameters of $\phi (1020)$.
Data from the same publication \cite {phipp} on $\phi K^+K^-$
contain a clear $f_0(980) \to KK$ peak, and the two sets of
data determine accurately the ratio $g^2(KK)/g^2(\pi \pi)$ of
couplings to $KK$ and $\pi \pi$.
An important detail is an error in units in \cite {phipp}:
$g^2(\pi \pi)$ is reported as 165 MeV; this should read
0.165 GeV$^2$.

The Breit-Wigner denominator for the $f_0(980)$ amplitude
is $[M^2 - s - i(g_1^2\rho _1 + g^2_2\rho _2)]$ and
$\rho _2$ has to be continued analytically below the
$KK$ threshold as $i\sqrt {4m^2_K/s - 1}$.
Without direct information on the $KK$ channel,
this term gets confused with $(M^2 - s)$ \cite {Baru}.
Any form factor applied to $g^2(KK)$ adds to the confusion.
One only has to glance at the Particle Data Tables \cite {PDG}
to see the large spread in parameters fitted to the $f_0(980)$
(and $a_0(980)$) in experiments having no direct information
on the branching ratio between $KK$ and $\pi \pi$.
Unfortunately, the PDG does not report the BES determination
of $g^2(KK)/g^2(\pi \pi )$, which is the best in the
published literature because of the availability of a
clear signal in $\phi KK$.

It was shown in \cite {Reconcil} that the BES parameters
for $f_0(980)$ are closely consistent with Kloe data \cite {Kloe}
on $\phi \to \pi ^0 \pi ^0 \gamma$ when one allows for
interference between $\sigma$ and $f_0$.
This paper determines $g^2(\sigma \to KK)/
g^2(\sigma \to \pi \pi ) = 0.6 \pm 0.1$.
BES data on $J/\Psi \to \omega K^+K^-$ confirm a large
value $\ge 0.6$ for this ratio \cite {WKK}.
The $\pi \pi $ full width at half-maximum (34 MeV) agrees well
with Cern-Munich data $30 \pm 10$ MeV \cite {Hyams}.
It also agrees closely with the full-width of the $f_0(980)$
signal in Crystal Barrel data $(\sim 46$ MeV).
The BES parametrisation will therefore be adopted in fitting
the AFS data.
Further checks from Belle, Babar and Cleo C will be very
welcome.

\section {Experimental tests}
\subsection {$J/\Psi \to \omega \pi\pi$}
Considerable detail needs to be given of fits to experimental
data, in order to pin down the disagreements with EU.
The prime conclusions which will emerge are that
(i) the $\pi \pi$ amplitude does not follow that of
elastic scattering, (ii) the magnitude of the $f_0(980)$
amplitude, relative to $\sigma$, is much smaller than
predicted by Eq. (14), regardless of analyticity
which also requires that their relative magnitudes should be
the same within $12\%$.

\begin{figure}[htb]
\begin {center}
\vskip -12mm
\epsfig {file=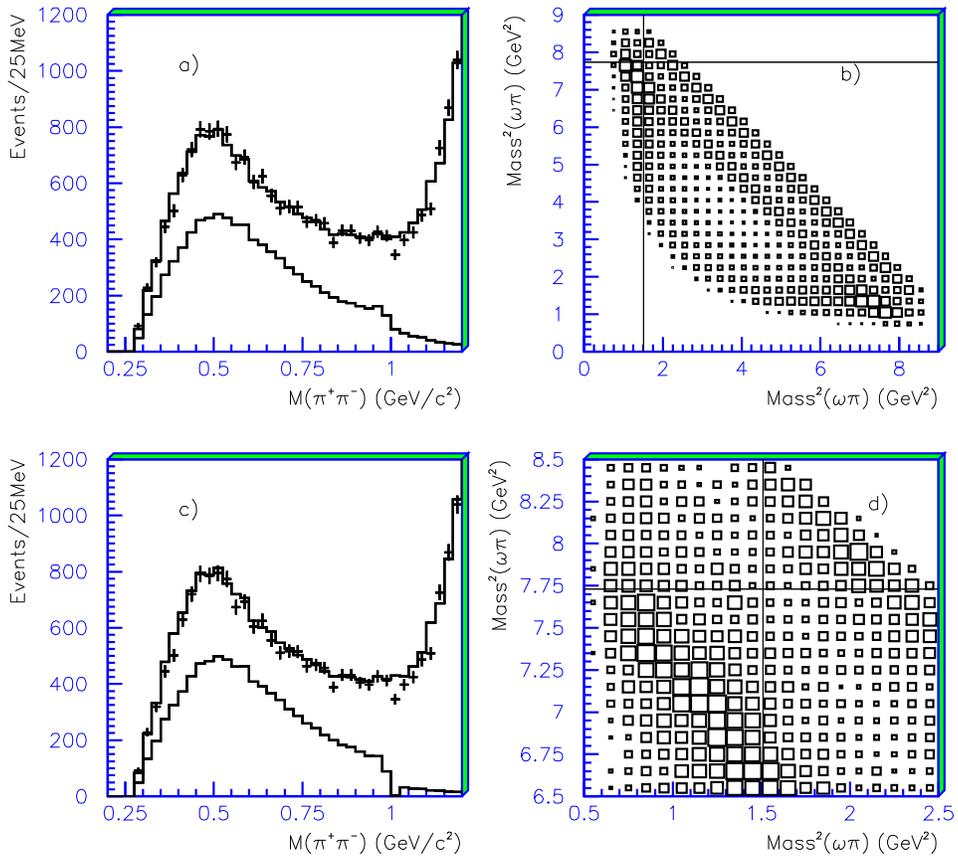,width=13cm}
\vskip -5mm
\caption {Fits to BES data for $J/\psi \to \omega \pi ^+\pi ^-$:
(a) $\pi \pi$ mass projection from the full Dalitz plot using the
$\sigma$  amplitude of Ref.~[13]; the lower histogram shows the S-wave
component; (b) Dalitz plot from data; (c) as (a) using Eq. (14)
for EU; (d) enlarged view of part of the Dalitz plot.}
\end {center}
\end{figure}

Fig. 4 displays features of BES data for
$J/\Psi \to \omega \pi \pi$ \cite {WPP}.
There is a clear peak in Fig. 4(a) at $\sim 0.5$ GeV due to the
$\sigma$ pole.
Its shape is cleanly separated from the $f_2(1270)$ contribution
up to 1.05 GeV, where the $f_2$ rapidly overtakes it.
The data include a slowly varying $14\%$ background
which is included in the fit.
There is also a well defined slowly varying component due to the
reflection of $b_1(1235)$.
Both are shown in the experimental  publication.

There are two amplitudes for production of $\sigma$ and $f_0(980)$,
with orbital angular momenta $L = 0$ and $2$ in the production
reaction.
The $L=2$ amplitude includes a centrifugal barrier for
production which optimises at a radius of 0.8 fm
(roughly as expected for convolution of wave functions of
$\sigma$, $\omega$ and $f_0$).
For the isobar model fit, different magnitudes are allowed for
$L = 0$ and 2, with complex coupling constants,
though it turns out that the $L=2$ contribution from
$f_0(980)$ is negligible.
In the EU fit the relative phases of $\sigma$ and $f_0$
components are constrained to be the same, and relative
magnitudes are constrained to be the same within $12\%$.
In Figs. 4(a) and (c), the lower histogram shows the $J^P=0^+$
intensity: from the isobar model fit in (a) and from EU in (c).
The upper histogram, close to data points, shows the result of
the full fit.
Unfortunately, the mass projection alone is not definitive, for
reasons explained shortly.
One low point in the mass projection just above 1 GeV hints at
a dip following the EU prediction.

However, there is a great deal more information contained in the
Dalitz plot of Fig. 4(b) and also in the correlation between the
$\omega$ decay plane and the $\omega \pi \pi$ production plane.
In the Dalitz plot, there are strong horizontal and vertical bands
due to $b_1(1235)$.
These bands interfere with $\sigma$ and $f_2(1270)$ (and $f_0(980)$
if present).
Note that the $b_1(1235)$ decays mostly through $S$-waves, so its
intensity would be almost constant across the Dalitz plot in the
absence of interferences.
Also note from the lower histogram of Fig. 4(a) that the $\sigma$
amplitude at 950 MeV is sizable.
Cross-hairs on Fig. 4(b) show where a $\pi \pi$ pair of mass 1 GeV
intersects the $b_1$ band.
If  $f_0(980)$ were present with the same magnitude as $\sigma$
and with the phase predicted by EU,
a large but narrow interference with $b_1$ cannot be avoided due to
$f_0(980)$, whatever the relative phase of $b_1(1235)$.
It should have a full width of $\sim 0.07$ GeV$^2$ along the
$b_1$ band: the line-width of $f_0(980) \to \pi \pi$.
There should also be a dip somewhere along the diagonal at
$m_{\pi \pi} = 1$ GeV.
Fig. 4(d) shows an enlarged view of this region.
There is no sign of these features.

The $\omega$ decay plays an important role.
The spin of the $\omega$ lies along the normal
to its decay plane; information on this decay plane is a key
ingredient in determining helicity amplitudes.
These are essential to  determine interferences between the
amplitudes for $\omega \sigma$,  $\omega f_2(1270)$ and
$b_1(1235)\pi$.
There are in principle five $f_2(1270)$ amplitudes corresponding
to production with $L = 0$ (one amplitude), $L = 2$ (three)
and $L = 4$ (one).
It turns out that the $L = 4$ amplitude and one of the three
$L = 2$ amplitudes are negligible.
There are large interferences between $\sigma$ and the remaining
three $f_2(1270)$ amplitudes and even larger interferences with
$b_1(1235)$.

With the EU hypothesis, the best fit to the Dalitz plot (upper
histogram) and the $\omega$ decay plane fills in the
predicted dip at $s_{\pi \pi} = 1$ GeV$^2$ with other interferences,
see Fig. 4(c); however, the price is a considerable deterioration
of log likelihood compared with the fit of Fig. 4(a).
The isobar model fit is better than EU by 259 in log likelihood.
There are two reasons.
Firstly, the narrow $f_0$ does not appear in interferences with
either $b_1$ or $f_2$.
The mass resolution of the BES data is 4 MeV in $\pi \pi$ at 950 MeV;
searching for the $f_0(980)$ with accurately known line-shape is
limited purely by statistics and there are $\sim 40$K events.
Secondly, in the fit based on EU, there are strong conflicts between
interference terms amongst $b_1(1235)\pi$, $\omega f_2(1270)$ and
$\omega \sigma$.
The data want slowly varying interferences between $b_1$ and the
broad $f_2$, instead of rapidly varying interferences of the narrow
$f_0$ with $b_1$ and/or $f_2$.

There are two extra fitting parameters for the isobar model fit:
i.e. two complex coupling constant for $\omega \sigma$ instead
of one. [The $f_0(980)\omega$ $L=2$ amplitude is negligible].
The definition of log likelihood is such that a change of 0.5
corresponds to a one standard deviation change in one variable.
For large statistics, $\chi^2 $ is approximately twice the change in
log likelihood.
So the fit of Fig. 4(a) is better than EU by 19.7 standard
deviations statistically.

It is important to remark that the BES publication gives a
second fit done independently.
This shows a $\pi \pi$ mass projection almost identical with
the scalar form factor.
This requires a very strong $f_0(980)$, in conflict with what
one can see by eye in Fig. 4(d).
However, this analysis made no use of the $\omega$ decay plane.
Without that information, it is impossible to disentangle
the magnitudes of the five $f_2(1270)$ amplitudes; the
determination of the magnitude and phase of the $f_0(980)$ signal
becomes very poor.
I have rechecked the analysis omitting information from the
$\omega$ decay plane.
No stable fit emerges with the $\pi \pi$ mass projection of the
second fit in the BES publication, and there is
no evidence for the presence of $f_0(980)$ at all in this fit.
The only explanation I can find of this second BES fit is that
the relative magnitudes of $f_0(980)$ and $\sigma$ have been
constrained to agree with the scalar form factor.
This is, of course, exactly what needs to be checked in the
present work.

\begin{figure}[htb]
\begin {center}
\vskip -12mm
\epsfig{file=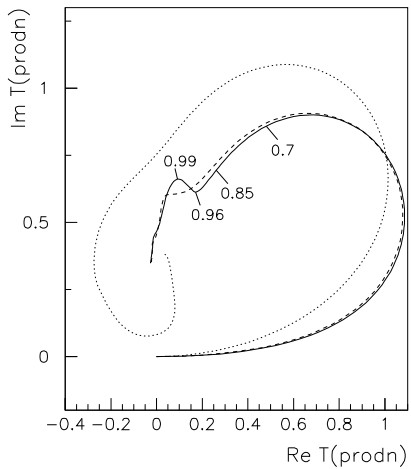,width=7.5cm}
\vskip -0.5cm
\caption {Argand diagram for $F\rho$;
masses are marked in GeV.
Full curve, fit; dashed curve, $\sigma$ alone; dotted curve,
EU prediction}.
\end {center}
\end{figure}

Fig. 5 shows the Argand diagram for $T(prodn) = F\rho$ from my
fit to $\sigma$ and $f_0$;
the factor $\rho$ is included so that the diagram goes to
zero at the $\pi \pi$ threshold.
The ratio of $f_0$ and $\sigma$ amplitudes is $0.12 \pm 0.06$
at 0.99 GeV and the $f_0$ lags the $\sigma$ by
$(42 \pm 20)^\circ$.
At the top of a large loop due to the $\sigma$, there is a
small structure due to $f_0(980)$.
For comparison purposes, the dashed curve shows the $\sigma$
amplitude alone and the dotted curve shows the EU prediction
normalised to the full curve at 0.47 GeV, the peak of the
$\sigma$.
The essential point is that the deep dip in this latter curve
near the $f_0$ mass is missing from the data.
There is one qualitative feature of Fig. 5 which will be important
later.
The amplitudes in a production process are free to add vectorially
in any way required by the data.
This situation is quite different from elastic scattering,
where the amplitude is constrained to the unitary circle.

It is now necessary to consider a number of systematic questions
which might blur the argument against EU.
Corresponding remarks will apply to fits to the other three
sets of data.

Firstly, could the disappearance of the predicted
$f_0(980)$ arise from effects of the $KK$ threshold?
The answer is no, because the $f_0$ appears clearly in elastic
scattering and the prediction of EU is that it should be almost
identical in production.
A precise cancellation of the magnitude and phase of the predicted $f_0$
would be needed. This is ruled out by the known couplings of both
$\sigma$ and $f_0$ to $KK$. The analytic continuation for the
$f_0(980)$ below the $KK$ threshold is accurately determined over the
small mass range concerned by BES data on $\phi \pi \pi$ and $\phi KK$.
In the $\sigma$ amplitude, the $KK$ inelasticity rises over a mass
scale of 200 MeV as shown in Fig. 11 of Ref. \cite {Reconcil}. Any
flexibility in the analytic continuation of the $KK$ term then has a
scale of $\sim 300$ MeV. Furthermore, it is closely constrained by the
fit to Kloe data on $\phi \to \gamma \pi ^0\pi ^0$. So this explanation
is highly implausible.

Secondly, could the BES data be fitted assuming
$J/\Psi\to\omega KK$, followed by $KK\to\pi\pi$?
This has been tested by adding $T_{12}$ fitted freely.
Its magnitude optimises at zero within errors.
It improves log likelihood only by 2.

\subsection {An objection of Aitchison}
Thirdly, Aitchison argues \cite {private}
that a further term $C$ might be added to the $P$-vector
due to dynamics of the production process.
A similar approach was used in fits by  Bowler et al. and Basdevant and Berger to the
$a_1(1260)$ to allow for the Deck effect \cite {Bowler}
\cite {AitB} \cite {Berger}.
An additional term $Ce^{i(\delta_A + \delta_B)}
( \cos \delta_A\cos \delta_B - \sin \delta_A \sin\delta_B)$
appears in $F\rho$.
If $C = i\alpha_A$, the second term can cancel the term
$iT_A T_B$ of EU, leaving $T_A$ and the first of the
additional terms, which is close to 0 since $\delta_A
\simeq 90^\circ$.
This removes almost all the structure due to $f_0(980)$.
However, this cancellation also leaves a term
$i \exp i(\delta_A + \delta_B) \cos \delta_A \cos \delta_B$
and when $\delta_B$ is small this becomes
$i \exp i\delta_A \cos \delta_A$, which is much larger
than the term $T_A$ itself if $C$ is a constant.
This is ruled out by the data, so it becomes necessary to
tailor the $s$-dependence of $C$ to reproduce the magnitude
of the $\pi \pi$ amplitude.

However, this is not the end of the story.
Phase information is also available.
In \cite {sigphase} it is shown that interference between
$\sigma$ and $b_1(1235)$ measures the phase of the $\sigma$
amplitude (plus any background term $C$) and requires it
to be the same as for elastic scattering down to 450 MeV
with an error of $5^\circ$ in the worst scenario, see
Fig. 2 of that paper.
It is worth mentioning that there is a similar result for the
$\kappa$ in \cite {kappa}.
Fig. 4(a) of that paper shows that the $\kappa$ phase from BES data on
$J/\Psi \to K^+\pi ^-K^-\pi^+$ agrees with the LASS effective
range formula for elastic scattering down to 750 MeV within
$3.5^\circ$.
Fig. 4(f) of the same paper shows a corresponding
agreement for E791 data on $D \to K\pi \pi$ within a similar
error.
These results rule out any background different from $\sigma$
and $\kappa$ with a magnitude larger than $12\%$ and a phase
difference above $\pm 5^\circ$.
They of course agree with Watson's original statement that
$D(s)$ of a single resonance should be universal.

There is no obvious source of the background
proposed by Aitchison, as there was in
$\cite {Bowler}$, where a Deck background was visible in the
data and made a $40\%$ contribution to cancelling the
$a_1(1260)$ amplitude.
A similar fortuitous cancellation with unidentified backgrounds
will be required also for all three of the following sets of
data.
In all cases, the additional background gives rise to a phase
variation different to strict EU.
If the background removes the $f_0$, only the $\sigma$ is left,
i.e. $T_\sigma$, which contributes only part of the phase of
elastic scattering; so the amplitude does not satisfy EU.

\begin{figure}[htb]
\begin {center}
\vskip -13mm
\epsfig{file=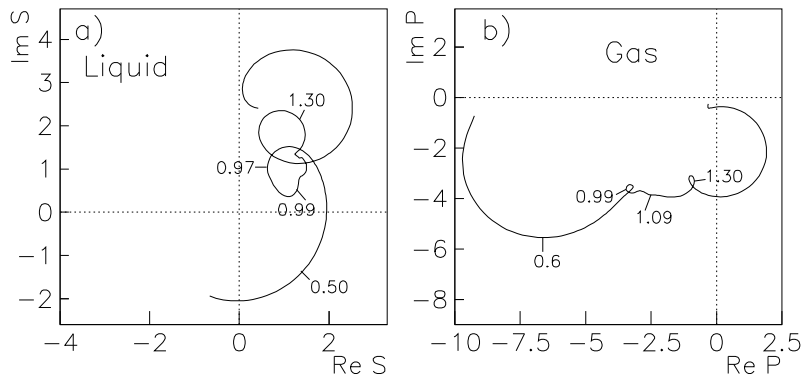,width=12cm}
\vskip -0.7cm
\caption {Argand diagrams for the $\pi \pi $ S-wave of Ref. [5]
for (a) $^1S_0$ annihilation, (b) $^3P_1$.}
\end {center}
\end{figure}

\subsection {$\bar pp \to 3\pi ^0$}
A fresh analysis of Crystal Barrel data on $\bar pp$ annihilation at
rest to $3\pi ^0$ has been completed recently \cite {f01370}.
Data are available with statistics of $\sim 700 $K  events (and
$0.5\%$ experimental backgrounds) in both liquid and gaseous
hydrogen;
these allow a good separation of $^1S_0$ and $^3P_1$
initial states.
Although the $f_0(980)$ appears clearly in $^1S_0$ annihilation,
the magnitude of the $\pi \pi$ S-wave amplitude does not go to
zero on the Argand diagram near 1 GeV, as EU predicts.
This Argand diagram is reproduced in Fig. 6(a).
Its magnitude is smallest at 0.98 GeV, but is still very distinct
from zero.
For $^3P_1$ annihilation, the fitted $f_0(980)$ amplitude, relative
to $\sigma$,  is much smaller, see Fig. 6(b).
The magnitude of the $\sigma$ amplitude is quite large near 1
GeV.

\begin{figure} [htb]
\begin {center}
\vskip -16mm
\epsfig{file=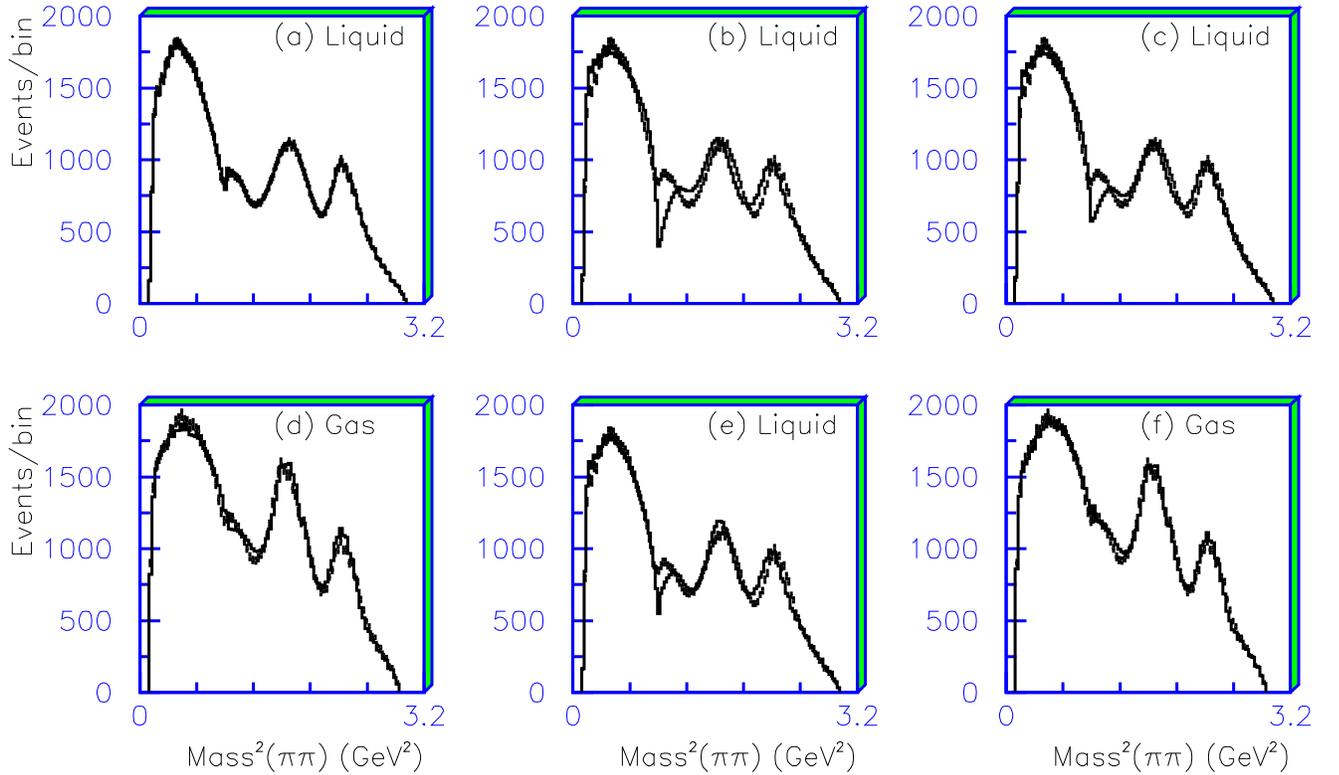,width=18cm}
\vskip -0.7cm
\caption {$\pi \pi $ mass projection of data and fits:
(a) from Ref. [5] in liquid hydrogen,
(b) with the fitted EU amplitude but other amplitudes untouched,
(c) after refitting all amplitudes,
(d) as (c) in gaseous hydrogen,
(e) and (f) with the $I=2$ S-wave included in the fit.}
\end {center}
\end{figure}
The question arises how robust these solutions are.
Could they `bend' to accomodate EU?
The published analysis requires some $s$-dependence of the
numerator fitted to the $\sigma$ amplitudes, of
the form $(1 + \Lambda s)$, with complex $\Lambda$.
Can extra flexibility reach agreement with EU?
A second point is that the analysis does not include the
repulsive $I = 2$ $\pi \pi$ S-wave.
Could this bring conclusions into line with EU?

The brief answer is definitely no, and will be presented
graphically.
Fig. 7(a) shows the $\pi \pi$ mass projection for data in liquid
from the current analysis; it fits the data points accurately.
In Fig. 7(b), the $\sigma + f_0$ combination of the isobar model
is replaced by the EU combination (with the constraint from
analyticity that magnitudes of $\sigma$ and $f_0(980)$
should be equal within $12\%$).
Initially, only the coupling constant of this combination is
refitted, leaving other amplitudes untouched;
this is for the purpose of illustrating the change required by
EU.
The $\sigma$ component is large and cannot change much; a deep
dip appears at 1 GeV because of the corresponding dip in the
elastic amplitude.

It is of course necessary to re-optimise all components.
The resulting mass projection is shown in Fig. 7(c) and is still
in severe disagreement with the data.
A measure of the disagreement may be obtained from $\chi^2$.
Here, it is necessary to point out that even the fit of
Fig. 7(a) has a $\chi^2$ larger than 1.
This is probably because of the enormous statistics and small,
slowly varying systematic errors in acceptance.
The procedure adopted here is to renormalise $\chi^2$ to 1 per
point for this fit and apply the same scaling factor to all
other fits.
The fit of Fig. 7(c) then has a renormalised $\chi^2$ of
40599 for 3500 bins; this is a 170 standard deviation
discrepancy, allowing for the reduction in the number of
fitted parameters by 2.

A much better fit is possible if relative
magnitudes of $\sigma$ and $f_0(980)$ are allowed to vary.
The phase of the S-wave amplitude in Fig. 6(a) is close to
EU, and only a 10 MeV shift is required for a
perfect fit.
This is consistent with the energy calibration
and resolution of the Crystal Barrel detector.
However, the fitted combination of amplitudes no longer
agrees with the crucial equation (14).

Fig. 7(f) shows the effect of including the $I=2$
S-wave amplitude, using the formulae of Section 4.1 of Ref.
\cite {f01370}.
There is only a rather small improvement, because the slow
$s$-dependence of this amplitude cannot fill the narrow dip at
1 GeV.
The renormalised $\chi^2$ falls to 34345, a discrepancy of
$\sim 130 \sigma$.
\begin{figure}[htb]
\vskip -12mm
\begin {center}
\epsfig{file=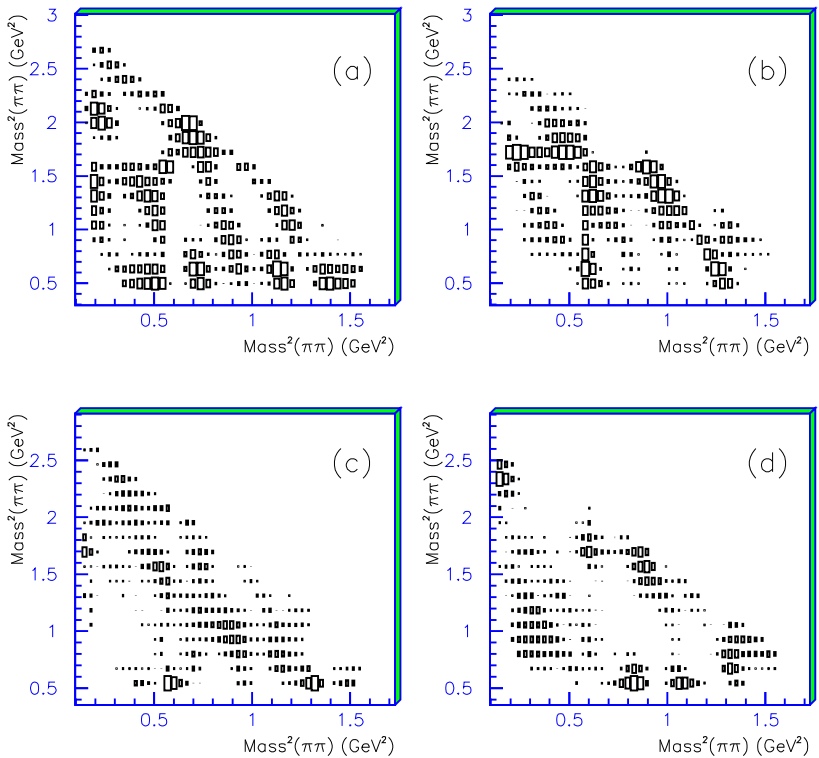,width=11.5cm}
\vskip -0.7cm
\caption {Discrepancies in $\chi^2$ over the Dalitz plot for (a) and (b)
liquid hydrogen, (c) and (d) gaseous hydrogen.
In (a) and (c) the fit is above the data and in (b) and (d)
below.}
\end {center}
\end{figure}

It is not just the $\pi \pi$ mass projection which governs $\chi^2$.
One should inspect discrepancies in $\chi^2$ all over the Dalitz
plot.
Fig. 8 makes such a comparison.
Panel (a) shows discrepancies in $\chi^2$ where the fit is above
the liquid data and (b) the discrepancies where the fit is low.
Panels (c) and (d) show results for data in gas.
One sees striking systematic discrepancies all over the Dalitz
plot, arising from interferences.
Such discrepancies are almost completely absent from the fit of
Fig. 7(a) using the isobar model.

There are two further points for discussion.
In the work shown on Figs. 7(e) and (f), the $I=2$ S-wave was fitted
without factoring out the Adler zero which occurs at $s \sim -0.41
m^2_\pi$.
If this is done (as for the $\sigma$ amplitude),
the broad $I=2$ amplitude gets confused with the $\sigma$
amplitude and leads to minor improvements all over the Dalitz
plot.
However,  none of these is distinctive enough to require
the $I=2$ amplitude definitively.

Secondly, could a more complicated polynomial than
$(1 + \Lambda s)$ multiply the $\sigma$ amplitude and give a
successful fit?
Extensive tests were made in \cite {f01370} with the objective
of improving the fits reported there.
If one chooses too free a polynomial, the fitted $^3P_1$ component
in gaseous hydrogen can fluctuate wildly from the 50\% predicted
from Stark mixing by Reifenrofer and Klempt \cite {RK}.
To avoid this, the fitted $^3P_1$ component is constrained
within the range $(50 \pm 7)\%$.
Unless the numerator of the EU amplitude is designed to include
a narrow dip at 1 GeV, no large improvement is observed.

One further observation from the Crystal Barrel data is worth
reporting.
Relative intensities of $f_2(1565)$ and $f_2(1270)$ are quite
different in $\bar pp$ annihilation to those in elastic scattering.
In $\bar pp$ data, the integrated intensities of these two
resonances are equal within $12\%$ after allowing for the (modest)
effects of centrifugal barriers for production
($L = 2$ for $^1S_0$ annihilation and $L=1$ for $^3P_1$).
However, in elastic scattering the fitted
$f_2(1565) \to \pi \pi$ width is a factor 4 smaller than that
of $f_2(1270) \to \pi \pi$.
This again disagrees with EU.

\section {Central production of $\pi \pi$}
Central production  of a $\pi \pi$ pair in $pp \to pp(\pi \pi)$
was fitted using EU by Au, Morgan and Pennington (AMP)
\cite {AMP}.
In data from the AFS experiment at the ISR \cite {AFS}, the two
final-state protons are produced with very small 4-momentum
transfers $t = -0.003$ GeV$^2$.
It is routinely assumed that the $\pi \pi$ pair is unaffected by
final state interactions with these protons, which are separated
from the central region by a gap in rapidity.

\begin{figure}[htb]
\vskip -15mm
\begin {center}
\epsfig {file=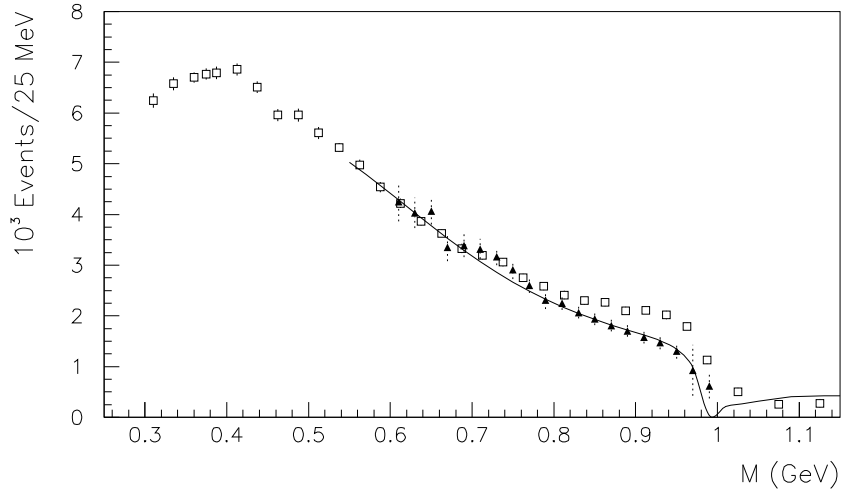,width=13.5cm}
\vskip -6mm
\caption {Open squares, AFS data; triangles, $\pi \pi$ cross
sections derived from phase shifts of Ochs [16];
curve, the elastic cross section with the numerator $N(s)$
replaced by a constant. }
\end {center}
\end{figure}

AMP draw attention to structure in the $\pi \pi$ mass spectum
similar to the dip in the $\pi \pi$ S-wave elastic cross
section.
Fig. 9 shows the AFS data as open squares.
There is a peak at low $\pi \pi$ mass, close to that of the
$\sigma$ pole in BES data, but not quite identical, for reasons
discussed shortly.

Triangles on Fig. 9 show $\pi \pi$ elastic cross sections derived from
Cern-Munich phase shifts; this is done by dividing the $\pi \pi$
amplitude of elastic scattering by $N(s)$, leaving only the term
$1/D(s)$.
The curve shows my fit to elastic data after dividing the $\pi \pi$
amplitude by $N(el)$.
The agreement between the curve and triangles demonstrates that
the parametrisation of the $\sigma$ reproduces Cern-Munich phase shifts.
Results are similar to the AFS data, but there is a distinct difference
in the vicinity of $f_0(980)$.
What is clearly evident is constructive interference between $f_0$
and $\sigma$ immediately below 989 MeV, where EU predicts a zero.
Even without fitting, one can see that EU will fail.

A detail is that, up to the $KK$ threshold, one piece of information
from each Cern-Munich moment is sufficient to determine phase
shifts for both S- and P-waves.
Above the $KK$ threshold, however, the separation of inelasticity
parameters $\eta$ and phase shifts $\delta$ cannot be made
without further assumptions.
Just above the $KK$ threshold, the solution of Ochs \cite {Ochs}
has $\eta=$0.6-0.7, whereas the BES line-shape for
$f_0(980)$ demands $\eta$ parameters dropping to $\sim 0.2$ at
1.01 GeV.
In view of this large discrepancy, predictions from Cern-Munich
phase shifts are not shown above the $KK$ threshold.

\begin{figure}[htb]
\vskip -12mm
\begin {center}
\epsfig {file=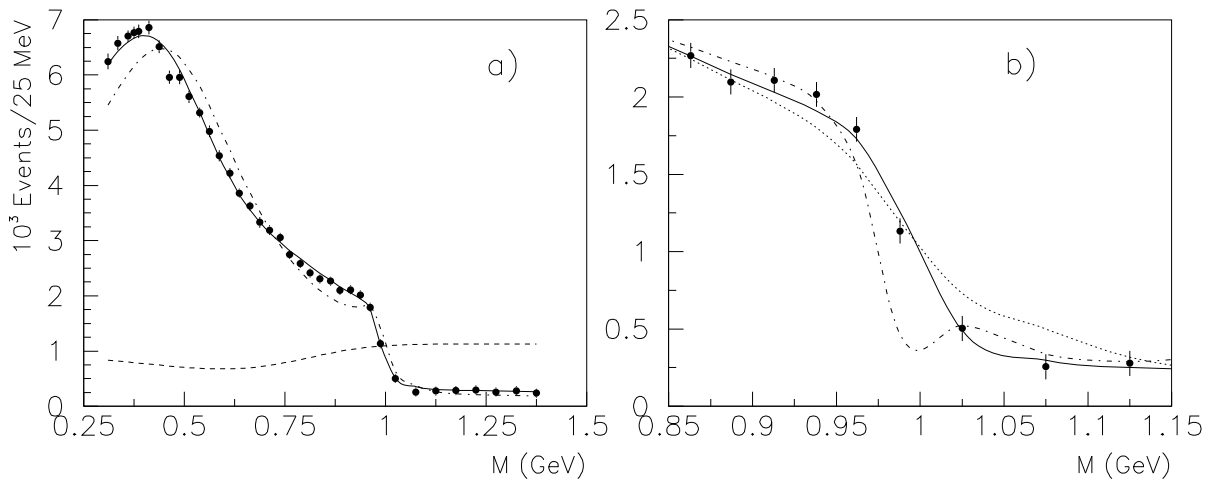,width=18cm}
\vskip -6mm
\caption {Fits to AFS data: (a) full curve isobar model,
chain curve using the $\sigma$ amplitude of Ref. [15],
dashed curve the scaling factor for the final fit.
(b) enlargement of the $f_0(980)$ region: full curve isobar
model, chain curve EU fit, dotted EU + freely fitted $I=2$ S-wave. }
\end {center}
\end{figure}

Since the work of Morgan and Pennington \cite {Morgan},
information on both $\sigma$ and $f_0(980)$ has improved
enormously.
The $\sigma$ amplitude is taken (within errors) from Ref.
\cite {DJphysG}, using fit (iii) given there.
The resulting cross section is shown by the chain curve of
Fig. 10(a).
Although this fit follows the general features of the data,
it is not accurate for low $\pi \pi$ masses.
Varying parameters of $f_0(980)$ within errors quoted by BES has
negligible effect on the quality of the fit.
Some additional flexibility is clearly needed in the numerator
$X(s)$ of the $\sigma$ amplitude.

There are two obvious sources.
Firstly, AMP point out that central production goes via
two intermediate Pomerons: $PP \to \pi \pi$, where
$\pi$ exchange is allowed.
This alone does not achieve a good fit.
The second explanation arises from
Regge factors in the production process.
AMP argue for an $s$-dependent factor $(m^2_\rho + s)$ in the
numerator arising from the leading Regge trajectory.
There may be contributions also from a daughter trajectory.

The full curve of Fig. 10(a) shows an isobar model fit using
a complex constant for $f_0(980)$, but without any
$f_0(1370)$.
It uses a scaling factor for the cross section
shown by the dashed curve.
This takes the form of a Gaussian dip:
$N(prodn) = 1 - A\exp -(s - s_0)^2$.
This scaling factor gives rise to a slow modulation of
the $\sigma$ amplitude over hundreds of MeV.
It has only a small effect on what is fitted to $f_0(980)$.
A detail is that a Gaussian mass resolution of 10 MeV, quoted by the
AFS collaboration \cite {AFS}, is folded into each point of the fit;
it is significant only near the $KK$ threshold.
The fit to $\pi \pi$ data has a $\chi^2$ of 35.2 for 34
degrees of freedom.

Consider now the $f_0(980)$ mass range.
The isobar model fit of Fig. 10(a) requires an $f_0$ magnitude
only $\sim 60\%$ of the EU prediction.
This fit requires constructive interference between $\sigma$
and $f_0$ below the $f_0$ mass and destructive interference
above.
It implies the $f_0$ phase is $(57 \pm 7)^\circ$ below the EU
prediction.
This immediately throws doubt on EU.

A fit using EU is conspicuously bad.
As an example, if the first 29 points up to 988 MeV are fitted
alone, $\chi^2 = 141.7$ if fitted by $\sigma$ and $f_0(980)$
only.
EU predicts an amplitude which is almost zero at 989 MeV,
whereas the data at 988 MeV are far above this.
If the EU fit is extended to 1300 MeV including the $f_0(1370)$
with the elasticity fitted in \cite {f01370},
$\chi^2$ is $>10^4$, because the $f_0(1370)$ contribution is
far too large.
Even if the $f_0(1370)$ is fitted freely in magnitude,
$\chi ^2 = 137$ for 37 points.
The fit, shown by the chain curve of Fig. 10(b) is
particularly bad close to the $KK$ threshold, where
the dip of elastic scattering is predicted by EU.
The fit to $KK$ data is also poor, with a $\chi^2$ of
18 for 5 points.

The dotted curve of Fig. 10(b) shows the effect of fitting
freely an additional contribution from the $I = 2$ S-waves:
$\chi ^2$ = 59.6.
This fails to cure the poor fit.
It makes almost no difference whether the $I = 2$ amplitude is
divided by a factor $(s - s_A)$ like the $\sigma$ amplitude.
The basic difficulty is that the slowly varying $I = 2$
amplitude cannot cure the rapid structure due to $f_0(980)$.
Furthermore, the relative magnitudes of the fitted $I = 2$
amplitude and the EU amplitude is 0.46, whereas it is only
0.18 for elastic scattering at 1 GeV.
Such a large $I = 2$ amplitude is implausible.

Obviously the problem with EU is that the
magnitude of the $f_0$ amplitude
needs to be smaller than for elastic scattering.
From analyticity, this also requires a difference in phase.
That is also clear from the absence of the predicted dip
at 989 MeV in the data.

If the phase of the $f_0(980)$ amplitude is constrained
to the EU value, but relative magnitudes of $\sigma$ and
$f_0$ are set free, $\chi^2 = 144$, which is bad.
Most of $\chi^2 $ comes from  points immediately
around the $KK$ threshold, showing that the data
reject also the phase variation of EU.

\subsection {Proposed Modification to EU}
At this point, one could argue that the mechanism of the
production reaction is unknown and might generate a phase
for $f_0(980)$ different to that of the $\sigma$.
This argument is not specific, though
the isobar model can fit the data well.
However, the usual argument for a different phase
for $f_0$ and $\sigma$ is multiple scattering of the pions
with spectator particles.
In AFS data, there is an empty rapidity gap isolating the central
region.
Remember also that Cern-Munich phase shifts are derived in the first
instance from data on $\pi p \to \pi \pi p$ at high momentum and
small momentum transfer, a similar configuration.

The conjecture of EU will now be replaced with an alternative ansatz.
The treatment of production data needs to be able to cope with the
case where one resonance amplitude is zero. EU does not, since
a universal phase equal to elastic scattering requires
a production amplitude $T^{(p)} \propto 1/D^\sigma(s)D^f_0(s)$.
The correct production amplitude should reduce to $T^\sigma$ when
the $f_0$ is absent. A small $f_0$ amplitude should produce a small
perturbation to $T_\sigma$.

Suppose the 2-body $\pi \pi$ amplitude is written
as
\begin {equation}
F\rho_1= \alpha T^{(p)} = \frac {\alpha '}{\sqrt {1 + \beta ^2}}
[T_A + \beta T_B e^{2i \Psi (s)}],
\end {equation}
where $\alpha '$ and $\beta$ are real.
This allows freedom in $\beta$ and includes
a phase $\Psi (s)$ which becomes the same as for elastic
scattering when $\beta \to 1$.
It is necessary to choose as A the state with the larger amplitude
on resonance, so that $\beta \le 1$.
If one could create this $\pi \pi$ system in `free space',
the appropriate 2-body unitarity relation below the $KK$
threshold
would be
\begin {equation}
{\rm Im} T^{(p)} = |T^{(p)}|^2.
\end {equation}

An alternative way of formulating the basic physics (with the same
outcome) is in terms of the Schwinger-Dyson equation.
Instead of the conventional relation
\begin {equation}
T_{prod} = V_{prod} + V_{prod} G T_{el},
\end {equation}
my conjecture is to replace this with
\begin {equation}
T_{prod} = V_{prod} + V_{prod} G T_{prod}.
\end {equation}
Here $V$ is the `potential' generating the final state and $G$ is
the propagator.

For the case of purely elastic scattering, a closed form for
$\Psi (s)$ of (19) may be derived by substituting $T_A$ and $T_B$
in the form  $(e^{2i\delta} - 1)/2i$ into (20).
After simple cancellations between left- and right-hand sides,
\begin {equation}
\sin (2\Psi + \delta_B - 2\delta_A) = \beta \sin \delta_B,
\end {equation} or
\begin {equation}
2\Psi = 2\delta_A - \delta_B + \sin ^{-1}(\beta \sin \delta_B).
\end {equation}
If $\beta \ne 1$, this is a different relation from purely
elastic scattering.

The improvement in the fit is dramatic.
Immediately an excellent fit to points below the $KK$
threshold is obtained with $\chi^2 = 28.4$ for 29 points and
24 degrees of freedom.
The term $\sin ^{-1}(\beta \sin \delta_B)$ in (24) differs
from $\delta_B$ by $54^\circ$.
This is just what is needed to produce the interference
between $\sigma$ and $f_0$ observed in the isobar model fit.

Above the $KK$ threshold, it is tempting to satisfy unitarity
by introducing a $K$-matrix.
However, the $K$-matrix depends on the assumption that the 2-body
system is confined to the unitary circle, but that is no longer the
case in a 3-body situation.

The fit may be extended above the inelastic threshold
by writing  $\pi \pi$ and $KK$ amplitudes as
\begin {eqnarray}
\nonumber
F_{\pi \pi } &=& \alpha [ T^\sigma_{11} + \gamma T^\sigma_{21}
+ (\beta T^f_{11} + \epsilon T^f_{21})e^{2i\Psi}]/\rho _1 \\
&=& \alpha \left[ T^\sigma_{11}
  \left(1 + \frac {\gamma g_2^\sigma r}{g_1^\sigma} \right)
  + T^f_{11} \left(\beta + \frac {\epsilon g_2^f r}{g_1^f}
\right) e^{2i\Psi} \right]\big/ \rho_1 \\\nonumber
F_{KK} &=& \alpha [T^\sigma_{12}  + \gamma T^\sigma_{22} +
(\beta T^f_{12} + \epsilon T^f_{22}) e^{2i\Psi '}]/\rho _2 \\
&=& \alpha
\left[ T^\sigma_{11}\frac {g^\sigma_2 }{g^\sigma_1 }
\left( \frac {1}{r} + \frac {\gamma g_2^\sigma }{g_1^\sigma}
\right) +
  T^f_{11} \frac {g^f_2}{g^f_1 }
  \left(\beta + \frac {\epsilon g_2^f } {g_1^f}
\right) e^{2i\Psi '} \right]\big/ \rho _1.
\end{eqnarray}
Eqs. (25) and (26) expose the explicit dependence of
$T_{12}$ on the ratio $r = \sqrt {\rho _2/\rho _1 }$;
the experimental group divides out the phase space $\rho_2$
and $\rho_1$ in the $\pi \pi$ and $KK$ channels.
As explained in Section 2.1, all $T$ have the numerator of
elastic scattering replaced by a constant.

My $T_{12}$ is defined so as to contain a factor
$\sqrt{\rho _1 \rho _2}$ and $T_{22}$ is defined to
contain a factor $\rho_2$.
With these definitions, the unitarity relations become
\begin {eqnarray}
{\rm Im}\, T_{11} &=& |T_{11}|^2 + |T_{12}|^2 \\
{\rm Im}\, T_{12} &=& T^*_{11}T_{12} + T^*_{12}T_{22} \\
{\rm Im}\, T_{22} &=& |T^*_{22}|^2 + T^*_{21}T_{12}.
\end {eqnarray}
Eqs. (25) and (26) satisfy these relations by construction,
except that $\Psi$ and $\Psi '$ need to be constrained
to obey (27)-(29).
Above the $KK$ threshold, this is done using freely fitted
$\Psi$ and $\Psi '$ for every data point and
introducing into $\chi^2$ a penalty function which applies
(27)-(29) with $3\%$ errors; in practise these constraints
are easily satisfied and discrepancies at the end of the fit
are below the $1\%$ level;
this is well below experimental errors.
In fact the $KK$ data are not very precise, leaving large
flexibility in $\Psi '$ above the $KK$ threshold.
In other words, the data are easy to fit above the $KK$
threshold, but highly definitive below it.

A detail is that $g_2$ needs to include form factors both
below and above the $KK$ threshold, such that it falls quite
rapidly on both sides of the threshold; formulae are given in
\cite {DJphysG}.

\begin{figure}[htb]
\vskip -16mm
\begin {center}
\epsfig {file=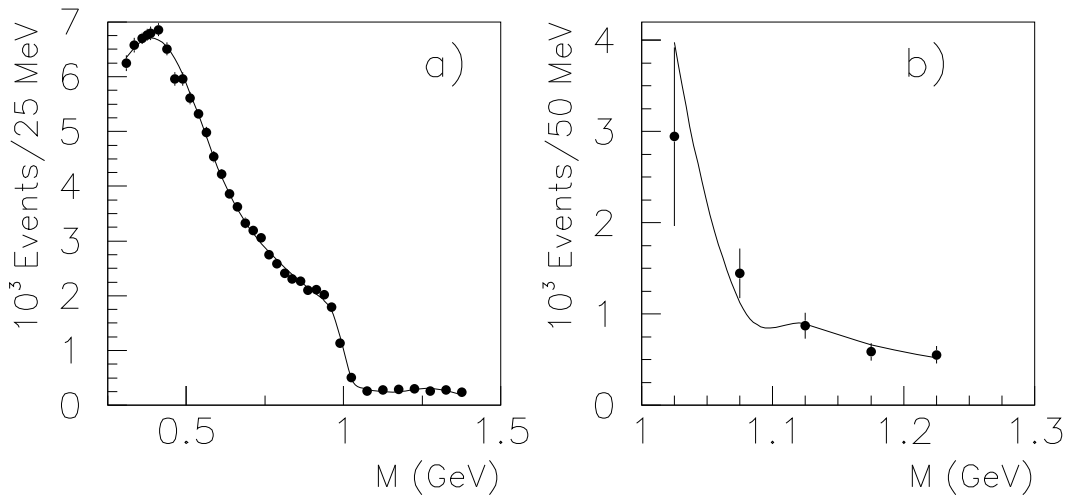,width=17cm}
\vskip -6mm
\caption {Fits to AFS data: (a) using the
revised form of 2-body unitarity,
(b) fit to $KK$ data. }
\end {center}
\end{figure}

The best fit with $\sigma$ and $f_0(980)$ alone has
$\chi^2 = 46.4$ for 37 points and 32 degrees of freedom.
The fit is good up to 1.1 GeV, but is inadequate for
$\pi\pi$ data near 1.3 GeV.
This may be cured straightforwardly by adding
a small $f_0(1370)$ component.
A technical detail is that the $f_0(1370)$ amplitude
is multiplied by a factor $\exp (2i\Psi'')$ and (24)
is iterated; the contribution of $f_0(1370)$ to $T_{12}$
and $T_{22}$ is negligible.
The resulting fit, shown on Fig. 11(a), has
$\beta = 0.59 \pm 0.06$ and the $f_0(1370)$ amplitude is 0.18
times that of the $\sigma$ amplitude at 1.3 GeV.
The $\chi^2$ for $\pi \pi$ data is 28.7 for 29 degrees of
freedom.
The Omega collaboration reports a significant contribution
from $f_0(1370)$ to their data on central production of
$\pi ^+\pi ^-$ \cite {Omegac}.
Their fitted mass and width agree closely with the line-shape
fitted to $f_0(1370)$ in Ref. \cite {f01370}.

Fig. 11(b) shows the fit to AFS $K^+K^-$ data.
A detail here is that the $K^+K^-$ data of AFS are scaled
up by a factor $4/3$ to allow for isospin Clebsch-Gordan
coefficients in $\pi ^+ \pi^-$ and $K^+K^-$ systems.
The $KK$ data constrain the coefficients of
$T_{12}$ and $T_{22}$ amplitudes.
The lowest AFS $KK$ point has small acceptance which may have
significant systematic uncertainty \cite {Carter}.

\begin{figure}[htb]
\begin {center}
\vskip -8mm
\epsfig {file=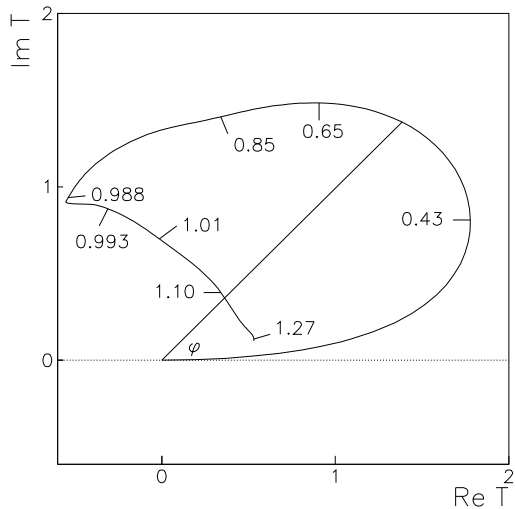,width=7.5cm}
\vskip -4mm
\caption {The Argand diagram of the amplitude fitting AFS data;
masses are marked in GeV. }
\end {center}
\end{figure}

Fig. 12 shows the Argand diagram for the fitted amplitude.
This illustrates the form of Eq. (20).
There is a geometrical relation between the imaginary
part of the amplitude and its modulus squared, but it is
a different relation to EU.
The $f_0(980)$ amplitude is smaller than that of the $\sigma$
and their relative phases are different to EU.
The same is true of Fig. 5, the Argand diagram fitting BES
data; there the $f_0(980)$ contribution is very small.

Let us  now return to equations (20) and (24) and review their
general features.
The last term of (24), $\sin ^{-1}(\beta \sin \delta_B) \to \delta_B$
as $\beta \to 1$ and $\delta _B \to 90^\circ$.
Furthermore it approaches this limit non-linearly as $\beta \to 1$.
The interesting point is that two-body elastic scattering emerges
as a limiting case of a more general relation.
Furthermore it has pathological properties as the relative magnitude
of A and B crosses 1 (or -1).
If B becomes larger than A, it is necessary to interchange the roles
of A and B.
As $\beta$ approaches 1 from below, $\Psi \to \delta_A$ and one
recovers the standard result of elastic scattering.
Further pathological cases arise if $\beta \to -1$ (impossible in
the 2-body elastic case).
Elastic scattering is in fact a very special case.
So EU can fail very badly as $\beta$ departs from 1.

As $\beta$ drops from 1, the phase $\Psi$ measured from the origin
of the Argand diagram soon changes by only a modest amount over
the $f_0$.
In this case, the isobar model becomes an excellent
approximation:
the $f_0$ has its usual dependence on $s$ through its
phase shift $\delta$ and
the $f_0$ amplitude adds vectorially to the  $\sigma$; the
isobar model can then fit a constant phase to
both quite successfully if the term
$\sin ^{-1}(\beta \sin \delta)$ is small.

In summary,
the modified form of EU suggested in (20) and (24) gives a much
better fit than EU.
The fit of Morgan and Pennington \cite {Morgan} using EU
requires an additional third-sheet pole at $M = 978 - i28 $ MeV.
This additional pole cannot be accomodated by BES
$\phi \pi \pi$ data, which require only a second-sheet pole at
$998-i17$ MeV and a broad third-sheet pole at $851-i418$ MeV.
These two poles have a natural explanation.
If the coupling of $f_0(980)$ to $KK$ is allowed to decrease
gradually to zero, leaving other parameters unchanged, the two
poles coalesce towards the same pole position $M = 968-i82$ MeV;
it is the effect of $\rho (KK)$ which moves the second-sheet
pole to the $KK$ threshold and moves the third-sheet pole away.
There is no pole in this mass range from the $\sigma$ amplitude.

A second remark is that all of $\sigma$, $\kappa$, $a_0(980)$
and $f_0(980)$ may be reproduced as a nonet by the model of
Rupp and van Beveren, where mesons couple to a quark loop
\cite {George}.
In this model, no additional pole like that of Morgan and
Pennington appears in the mass range close to the $KK$ threshold.
These two results suggest that the additional
pole is a consequence of the constraint of EU.

The relation (20) is a new conjecture.
Are there forseeable snags?
The $\sigma$ and $f_0$ may mix, and this mixing could be
different in elastic scattering and production.
This mixing would alter the apparent width of $f_0(980)$
and could induce an additional phase change relative to
$\sigma$.
Presently there is no indication of any need for mixing.
Such mixing will be absent if $\sigma$ and $f_0(980)$ have
strictly orthogonal wave functions, as is plausible for
members of the same nonet.
On Fig. 2, the $\sigma$ is definitely visible in $\phi \pi \pi$
data.
It would not be surprising if $\phi \pi \pi$ and $\omega \pi \pi$
channels filter out orthogonal combinations of $\sigma$ and $f_0$.
From Fig. 2, one can estimate the relative intensities of
$\sigma$ and $f_0$.
It is necessary to allow for the mass resolution, since the
$f_0$ amplitude falls extremely rapidly from its peak at
989 MeV, particularly above the $KK$ threshold.
Doing this, the intensity of $\sigma$ is $4\%$ of $f_0$ at the
peak, i.e. $20\%$ in amplitde.
This is marginally higher than the $f_0$ signal fitted to
$\omega \pi \pi$ but within the error, supporting the idea
of orthogonal amplitudes.

\section {How to fit elastic data above the $KK$ threshold}
Many authors use the $K$-matrix to satisfy unitarity for
2-body scattering, e.g. the coupled channels $\pi \pi$, $KK$,
$\eta \eta$, $\eta \eta '$ and $4\pi$.
The popular approach is to add $K$-matrices of all resonances
appearing in one partial wave.
However, if resonances overlap, as $\sigma$ and $f_0(980)$ do,
the $K$-matrix poles occur at masses where combined phase
shifts happen to go through $90^\circ$, $270^\circ$, etc.,
i.e. at $\sim 750$ and 1200 MeV.
Firstly, an expansion in terms of these poles is problematical
for $f_0(980)$ unless other factors or high powers of $s$ are
included.
Secondly, the relation between $K$-matrix poles and $T$-matrix
poles is obscure.
Any one $T$-matrix pole is built up from all $K$-matrix poles;
the converse is obvious.
The prescription that $S$-matrices multiply below the inelastic
threshold does not appear naturally, but has to be enforced by
fitting data.

An attractive alternative can be constructed following the
spirit of Aitchison's approach (for 2-body scattering).
Suppose one combines two resonances
according to the prescription
\begin {equation}
K_{ij}(total) = \frac {(K_A + K_B)_{ij}}{1 -
0.5\rho_i\rho_j(K_AK_B + K_BK_A)_{ij}}.
\end {equation}
Below the $KK$ threshold, this automatically
gives the result that phase shifts add. [Further resonances may be
combined by iterating this prescription]. A nice feature of (30) is
that one can write
\begin {equation}
K_{ij} = \frac {g_ig_j}{M^2 - s},
\end {equation}
using the same mass $M$ as the usual Breit-Wigner denominator.
A second attractive feature of (31) is that the amplitude
continues naturally through the $KK$ threshold, because of
the factor $\rho_i\rho_j$ in the denominator.

My own approach in several papers, \cite {Reconcil, f01370}
has been close to this.
All diagonal elements of $S$-matrices are multiplied, as proposed
here.
Magnitudes of off-diagonal elements of
the $S$-matrix need to be calculated from unitarity relations.
For example, for a 3-channel system:
\begin {equation}
|S_{12}|^2 = (1/2) (1 + |S_{33}|^2 - |S_{11}|^2 - |S_{22}|^2).
\end {equation}
The phase of these off-diagonal elements has been fitted
empirically, whereas (30) would predict these phases.
This approach successfully fits elastic data, $\pi \pi \to KK$ and
$\eta \eta$ with one proviso: a good fit requires inclusion of mixing
between $\sigma$, $f_0(1370)$ and $f_0(1500)$ \cite {f01370}, using the
formulae of Anisovich, Anisovich and Sarantsev \cite {AAS}; these
formulae are the modern equivalent of the Breit-Rabi equation of
molecular spectroscopy, generalised to include resonance widths.

\section {Conclusions}
Crystal Barrel data have $f_0(980)$ and $\sigma$ components
with relative magnitudes seriously different to those predicted by EU
plus analyticity.
Furthermore, they are different in $^3S_1$ and $^3P_1$
annihilation.
The BES data for $J/\Psi \to \omega \pi ^+\pi ^-$ do not
reproduce the deep dip of elastic scattering at 989 MeV.
AFS data likewise do not contain the same dip at this mass.
All these results are in conflict with equation (14), which is
a direct consequence of EU.
This shows unambiguously that there must be a major flaw in the
hypothesis of EU.

Experimentalists have dealt with this problem by using
complex coupling constants for each resonance.
However, as emphasised in subsection 2.1, the imaginary part of
the coupling constant introduces into the numerator of the
amplitude an $s$-dependent phase variation which alters the
universal phase coming from the denominator $[1 - i \rho K(s)]$.
This destroys the original idea of a universal phase.
One might as well fit directly in terms of the $T$-matrix of each
individual resonance, along the lines outlined in Section 5.
The form of the $K$-matrix suggested there would eliminate
differences between $K$-matrix and $T$-matrix poles, making
interpretation of results more direct.

Not all experimentalists adopt the P-vector approach.
Some fit directly in terms of individual $T$-matrices for each
resonance, including sequential decays from one resonance to
a daughter with different complex coupling constants for
each decay mode.
Ascoli and Wyld \cite {Ascoli} and Schult and Wyld
\cite {Schult} consider a multiple scattering series of the type
$R \to (12)3 \to 1(23)$, etc, where $R$ is a 3-body resonance and
brackets indicate resonances in two-body sub-systems; this is a
unitarity effect of a different form to that considered here.

In view of the failure of EU in the 4 cases considered here, each new
set of data should be inspected on its merits.

Let us examine ways of trying to save EU.
Firstly, it is possible that unspecified backgrounds can be added to
the P-vector so as to side-step the conflict.
However, analyticity independently limits relative magnitudes
of $f_0(980)$ and $\sigma$ within $12\%$.
Experimental determinations of $\sigma$ and $\kappa$ phases
in \cite {sigphase} and \cite {kappa}
constrain phases within $\pm 5^\circ$.
The probability that unspecified backgrounds can evade Eq. (14) to
this accuracy in four sets of data is vanishingly small.
Furthermore, if such backgrounds are introduced, EU loses
any predictive power.

Secondly, Crystal Barrel data and AFS data cannot be fitted with EU
whether or not the $I=2$ S-wave amplitude is included.
So this is not a satisfactory escape route.

A third likely possibility, applicable to the first three
sets of data, is that pions from sigma and/or $f_0(980)$ rescatter
from the spectator particle, leading to a breakdown of EU.
Aitchison himself pointed this out.
Today, we known that such graphs have magnitudes  typically $25\%$
of those of the parent processes before the rescattering.
This is sufficient to introduce large phase changes in
some cases, but not all.

For all of these three sets of data, the isobar model provides an
excellent fit.
The production amplitude is then written $F = \alpha_A
N_A(el)/D_A + \alpha_B N_B(el)/D_B$ with complex $\alpha_A$ and
$\alpha_B$.
In this form, no vestige remains of the constraint that
$S$-matrices must multiply as in elastic scattering.
Relative magnitudes of $\alpha_A$ and $\alpha_B$ can arise from
matrix elements coupling the initial state to each resonance.
For $J/\Psi \to \omega \pi \pi$ and $^3P_1~\bar pp \to 3\pi ^0$,
the $f_0$ component is so small that one cannot tell whether
the phase alone follows EU or not.
However, for $^1S_0~\bar pp \to 3\pi ^0$, the $f_0$ signal is large
enough to rule out this possibility.

The fourth point is that one would still expect EU to work for AFS
data, but it does not.
An excellent fit may be obtained by replacing the unitarity relation
$\rm {Im}\, T^{(p)} = T^{(p)}T_{el}^*$ by the new relation
\begin {equation}
\rm {Im}\, T^{(p)} = |T^{(p)}|^2.
\end {equation}
This corresponds to the relation
\begin {equation}
{\rm Im}\, F = F T^{*(p)},
\end {equation}
rather than the commonly used form $ {\rm Im}\, F = F T^*_{el}$.

Equations (20) and (24) have pathological behaviour in the vicinity
of the 2-body elastic limit $\beta = 1$.
One can now see the basic problem of EU.
It attempts to impose on the 3-body system a very special behaviour
which is narrowly restricted to 2-body scattering.
Away from the special case $\beta = 1$, the isobar model works
successfully.

There are two points about the new unitarity relations.
Firstly, it was argued in Section 2.1 that a universal phase in the
denominator of the amplitude also requires, via analyticity, almost
universal magnitudes;
the word `almost' covers the possibility that there may be slowly
varying form factors or centrifugal barrier factors in production
reactions without corresponding changes to $D(s)$.
If relative magnitudes of resonances differ by large amounts between
2-body scattering and production, their relative phases {\it must} also
change.

Secondly, the new unitarity relation (33) succeeds
quantitatively in accounting for the observed relative phase
between $\sigma$ and $f_0(980)$ in central production.
That is a non-trivial result.
The fit to AFS data then requires only two poles in the vicinity of
$f_0(980)$, in agreement with the BES line-shape (as does the
isobar model).
EU requires an extra pole for which there is no obvious explanation.
The form of this new unitarity relation is illustrated by the Argand
diagrams of Figs. 5 and 12.
In both, the $f_0(980)$ amplitude is small or fairly small and
so is $\sin ^{-1}(\beta \sin \delta_B)$ of Eq. (19).
As a result, the phase $\Psi$  measured from the origin of the
Argand diagram changes rather little over the $f_0$.
This is an extra source of phases appearing in the
isobar model, and has not been appreciated before.

However, one then needs to ask whether this new relation can be
used universally in the isobar model.
Does it, for example, correctly describe the relative phases
of $\sigma$ and $f_0$ in $^1S_0 \to 3\pi ^0$ data and in
$J/\Psi \to \phi \pi \pi$?
The answer is no.
For these two reactions, the agreement between data and equations
(20) and (24) improves substantially over EU. However, there are
still discrepancies with the new unitarity relation of
$20-30^\circ$,
which is still significant.
It seems likely that rescattering
of pions from the spectator introduces some additional phases.

The new unitarity relation needs to be tested elsewhere.
A possible testing ground is in Kloe data on $\phi \to \gamma \pi ^0
\pi ^0$, where both $\sigma$ and $f_0$ may contribute, but the
decay is electromagnetic; the small amplitude from
$\phi \to \rho \pi^0$ introduces a perturbation of only $4\%$ in
amplitude and only in well defined parts of the Dalitz plot.

The remedy which succeeds well in fitting nearly all data is the
isobar model, where both magnitudes and phases of resonances
are both fitted freely.
It needs to be emphasised that experimental groups have adopted
the flexibility needed to fit existing data, so their results
are essentially sound and are not in question.
The hypothesis of EU has mostly been adopted by theorists for
making predictions.
Those predictions now need to be viewed with suspicion.
Although the scalar form factor is well
determined for elastic scattering, it is dangerous to assume
that this form factor is universal and can predict production
processes.

\section*{Acknowledgements}
I am greatly indebted to I.J.R. Aitchison for extensive and
illuminating comments.
I am also grateful to Prof. G. Rupp and Prof. E. van Beveren
for extensive discussions.

\end {document}